\documentclass[aps,epsfig,prb,unsortedaddress,twocolumn]{revtex4}
\usepackage{amsmath}
\usepackage{graphicx}

\def\nn{\nonumber}

\begin{document}

\title{Determination of intrinsic lifetime of edge magnetoplasmons}

\author{Ken-ichi Sasaki}
\email{sasaki.kenichi@lab.ntt.co.jp}
\affiliation{NTT Basic Research Laboratories, NTT Corporation,
3-1 Morinosato Wakamiya, Atsugi, Kanagawa 243-0198, Japan}

\author{Shuichi Murakami}
\affiliation{Department of Physics, Tokyo Institute of Technology,
2-12-1 Ookayama, Meguro, Tokyo 152-8551, Japan}

\author{Yasuhiro Tokura}
\affiliation{NTT Basic Research Laboratories, NTT Corporation,
3-1 Morinosato Wakamiya, Atsugi, Kanagawa 243-0198, Japan}
\affiliation{Faculty of Pure and Applied Sciences,
University of Tsukuba, Tsukuba, Ibaraki 305-8571, Japan}

\author{Hideki Yamamoto}
\affiliation{NTT Basic Research Laboratories, NTT Corporation,
3-1 Morinosato Wakamiya, Atsugi, Kanagawa 243-0198, Japan}

\date{\today}
 
\begin{abstract}
 It is known that peculiar plasmons whose frequencies are purely imaginary exist
 in the interior of a two-dimensional electronic system described by the Drude model.
 We show that when an external magnetic
 field is applied to the system,
 these bulk plasmons are still non-oscillating and are isolated from 
 the magnetoplasmons by the energy gap of the cyclotron frequency. 
 These are mainly in a transverse magnetic mode and can
 combine with a transverse electronic mode locally at an edge of the
 system to form edge magnetoplasmons.
 With this observation, we reveal the intrinsic long lifetime of edge
 magnetoplasmons for the first time.
\end{abstract}

\maketitle

Many types of intriguing phenomena can emerge at the edge of a material that are
invisible or hiding in the interior for some reason.
Edge magnetoplasmon is such an example; 
it is a gapless collective excitation that appears at the edge of a
two-dimensional electron gas (2DEG) under the application of an external
magnetic
field.~\cite{Allen1983,Glattli1985,Grodnensky1991,Ashoori1992,Tonouchi1994,Yan2012,Petkovic2013,Kumada2013}
The properties unique to the edge magnetoplasmons, such as the localization length and dispersion relation, 
were calculated by Volkov and Mikhailov.~\cite{volkov88}
They succeeded in solving an integral equation of the electric
potential using the Wiener-Hopf method.~\cite{noble1958,fetter1985}
Meanwhile, an internal magnetic field, which is coupled to the potential
through Maxwell equations, is neglected, and this simplification
prevents the lifetime of the edge magnetoplasmon
($\tau^*$ in Ref.~\onlinecite{volkov88}) from being determined and
also obscures the magnetic configurations of the excitations.
The fact that the localization length and dispersion relation are
dependent on the lifetime makes it difficult to analyze experimental results.

In this paper, we determine the intrinsic lifetime of the edge
magnetoplasmon.
Our analyses are based on two observations. 
The first is that 
there is a purely relaxational state with a very long lifetime in
the interior of a 2DEG.
The second is 
that the state acquires a non-zero real part of the frequency through
localization and starts to propagate.
By showing that the properties of the localized state are consistent
with those of the edge magnetoplasmons, we identified
the purely relaxational state as the bulk counterpart of the edge
magnetoplasmons and determined the lifetime of the edge magnetoplasmons.
Our results show that the internal magnetic field normal to the layer is
strongly suppressed in the interior, which partly justifies the assumption used in
the past and may lead us to a more complete description of the edge
magnetoplasmons.

\begin{figure}[htbp]
 \begin{center}
  \includegraphics[scale=0.7]{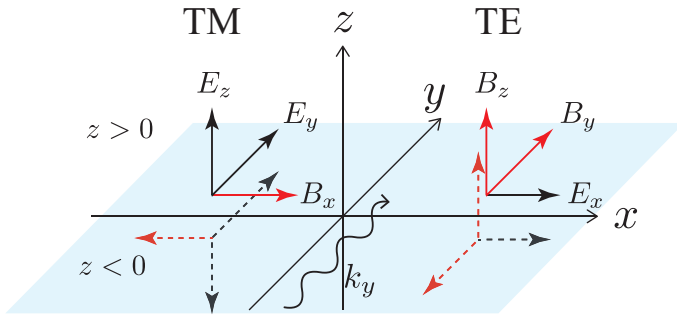}
 \end{center}
 \caption{(color online) 
 The electromagnetic fields (${\bf E}$ and ${\bf B}$) of the transverse
 magnetic (TM) and transverse electric (TE) modes are shown for
 $z>0$ (solid arrows) and $z<0$ (dashed). 
 The discontinuities of $B_x$ ($E_z$) and $B_y$ at $z=0$ 
 are relevant to the TM and TE modes, respectively.
 In contrast to $E_z$,
 $B_z$ cannot be discontinuous at $z=0$ because of the absence of a
 magnetic monopole.
 A 2DEG layer at $z=0$ is expressed by a transparent sheet.
}
 \label{fig:TMTE}
\end{figure}

We begin by reviewing a mathematical treatment of plasmons.
The fact that magnetic fields are discontinuous at a 2DEG layer plays a central role in
the formation of localized surface plasmons.~\cite{nakayama74,Chiu1974} 
The plasmons are classified into transverse magnetic
(TM) and transverse electric (TE) modes with respect to their eigenvectors,
as shown in Fig.~\ref{fig:TMTE}.
The electric fields of these eigenmodes are written in
terms of the localization length in the direction normal to the layer
$\alpha^{-1}$, in-plane wavevector $k_y$, and angular frequency $\omega$ as
\begin{align}
 \begin{split}
  & E_x = E^{\rm TE}_{x0} e^{i(k_y y -\omega t)} e^{-\alpha|z|}, \\
  & E_y = E^{\rm TM}_{y0} e^{i(k_y y -\omega t)} e^{-\alpha|z|}, \\
  & E_z = E^{\rm TM}_{z0} e^{i(k_y y -\omega t)} e^{-\alpha|z|},
  \label{eq:Efield1}
 \end{split}
\end{align}
where $E^{\rm TE}_{x0}$ ($E^{\rm TM}_{y0}$ and $E^{\rm TM}_{z0}$)
is the amplitude of the TE (TM) mode.
Note that the $E_i$ values in Eq.~(\ref{eq:Efield1}) are proportional to
$e^{-\alpha z}$ ($e^{+\alpha z}$) for $z>0$ ($z<0$).
Note also that for $z<0$ we need to replace $E_{z}$ with $-E_z$ 
because Gauss's law for free space, $\nabla \cdot {\bf E}=0$,
must be satisfied for both $z>0$ and $z<0$.
The magnetic fields are obtained from Eq.~(\ref{eq:Efield1}) using 
Faraday's law, $\nabla \times {\bf E} = -\partial {\bf B}/\partial t$.
For $z>0$, we have
\begin{align}
 \begin{split}
  & i\omega B_x = ik_y E_z + \alpha E_y, \\
  & i\omega B_y = -\alpha E_x, \\
  & i\omega B_z = - ik_y E_x.
 \end{split}
 \label{eq:Bfield1}
\end{align}
For $z<0$, we replace $\alpha$ and $E_z$ on the right-hand side of 
Eq.~(\ref{eq:Bfield1}) with $-\alpha$ and $-E_z$, respectively.
As a result, $B_{x}$ and $B_y$ are discontinuous at $z=0$ 
as shown in Fig.~\ref{fig:TMTE}.
By applying Stokes' theorem to Amp\'ere's circuital law of the
Maxwell equations, 
$c^2 \nabla \times {\bf B} = \varepsilon \partial {\bf E}/\partial t +
{\bf j}/\epsilon_0$ where ${\bf j}=(j_x,j_y,0)\delta(z)$ is the
electronic current flowing in a layer, 
we find that  
the discontinuity of $B_x$ is related to $j_y$ as
$c^2 (B_x(z=0+)-B_x(z=0-))=j_y/\epsilon_0$. 
Because of Ohm's law, ${\bf j}$ (on the right-hand side) is
proportional to the in-plane electric fields as $j_i =
\sigma_{ix}(\omega)E_x|_{z=0} + \sigma_{iy}(\omega)E_y|_{z=0}$ with the coefficients
of the dynamical conductivity tensor $\sigma_{ij}(\omega)$ given below.~\footnote{We assume
that $\sigma_{xx}(\omega)=\sigma_{yy}(\omega)$ and $\sigma_{xy}(\omega)=-\sigma_{yx}(\omega)$.}
By applying a similar argument for $B_y$, 
we obtain the equations for the amplitudes,
\begin{align}
 \begin{pmatrix}
  \frac{2\alpha}{i\omega \mu_0} - \sigma_{xx}(\omega)
  & - \sigma_{xy}(\omega) \cr
  \sigma_{xy}(\omega) & 
  \frac{2 i\omega \epsilon}{\alpha} - \sigma_{xx}(\omega)
 \end{pmatrix}
 \begin{pmatrix}
  E^{\rm TE}_{x0} \cr E^{\rm TM}_{y0}
 \end{pmatrix}
 = 0.
 \label{eq:mat1}
\end{align}
Here, $\mu_0$ ($\epsilon_0$) is the permeability (permittivity) of free space, 
\begin{align}
 \alpha=\sqrt{k_y^2 - \varepsilon \frac{\omega^2}{c^2}},
 \label{eq:alpha}
\end{align}
and $\varepsilon \equiv \epsilon/\epsilon_0$ is the relative
permittivity of the surrounding material.
We assume that $\varepsilon$ is a frequency-independent constant
throughout this paper.
A detailed derivation of Eqs.~(\ref{eq:mat1}) and (\ref{eq:alpha}) is
given in Appendix.

We adopt the Drude model to calculate $\sigma_{ij}(\omega)$.
The model describes the motion of an electron accelerated by the
electric fields in an applied static magnetic field, ${\bf B}_{a}$.
This motion is governed by the classical equation of motion:
$m(d{\bf v}/dt+{\bf v}/\tau)=-e({\bf E}+{\bf v}\times {\bf B}_a)$, 
where $m$ is the effective mass of the electron, $\tau$ is the
relaxation time, and ${\bf v}$ is the velocity.
The solution of the equation gives, 
with the definition of the current ${\bf j}\equiv -e n {\bf v}$
($n$ is the carrier density), the conductivities of the Drude model as
\begin{align}
 \begin{split}
  & \sigma_{xx}(\omega) = \frac{(1-i\omega \tau)\sigma_0}{(1-i\omega \tau)^2 +
  (\omega_c \tau)^2}, \\
  & \sigma_{xy}(\omega) = -\frac{(\omega_c \tau)\sigma_0}{(1-i\omega \tau)^2 +
  (\omega_c \tau)^2},
 \end{split}
 \label{eq:drude}
\end{align}
where $\sigma_0=ne^2\tau/m$ is the static conductivity and $\omega_c =
eB_{az}/m$ is the cyclotron frequency.
The frequency and eigenvector of the surface plasmons are determined
from Eq.~(\ref{eq:mat1}) with Eq.~(\ref{eq:drude}).

It has been shown by Fal'ko and Khmel'nitskii that plasmons 
whose frequencies have no real part exist when $\omega_c=0$.~\cite{falko1989}
The off-diagonal terms of Eq.~(\ref{eq:mat1}) disappear, 
so that the TM and TE modes are decoupled completely.
The TM mode satisfies the quadratic equation with respect to $\omega$,
\begin{align}
 \frac{2 i\omega \epsilon}{\alpha} -\frac{\sigma_0}{1-i\omega\tau}=0.
 \label{eq:d1}
\end{align}
In particular, when $\tau \ll \epsilon/2 \sigma_0 \alpha$,
we obtain two roots corresponding to a long-lived mode
with $\omega \simeq -i \sigma_0|k_y|/2\epsilon$
and a short-lived mode with $\omega \simeq -i/\tau$.~\footnote{Here, 
we are interested in the low-energy plasmons for which $\alpha$ can be
well approximated by $|k_y|$.}
The frequencies of these modes have a zero real part
and are expressed as $\omega = -i\delta$ with a positive real number $\delta$.
The time evolution exhibits an exponential decay (overdamped oscillation),
$e^{-i\omega t} = e^{-\delta t}$, in other words,
they are non-oscillating and purely relaxational states.
For the short-lived mode, the lifetime is identical to the relaxation
time of the electron, suggesting that the mode is controlled by the
electron's motion.
The lifetime of the long-lived mode is inversely proportional to $\tau$ 
since $\sigma_0$ is proportional to $\tau$ and is enhanced in the
long-wavelength limit $|k_y|\to 0$.~\footnote{
Note that the $k_y$ dependence of $\delta$ causes a spatial change from
the initial configuration of the electromagnetic fields, for example in the diffusion.}
These purely relaxational states are distinct from the mode extensively
discussed in the literature that appears in the collisionless limit
$\tau \gg \epsilon/2 \sigma_0 \alpha$.~\cite{stern67,chaplik72,Wunsch2006,hwang07,Sasaki2014}
The mode oscillates with the frequency, 
\begin{align}
 \omega \simeq \pm \omega_p -\frac{i}{2\tau},
\end{align}
where
\begin{align}
 \omega_p \equiv \sqrt{\frac{\sigma_0 \alpha}{2\epsilon \tau}}.
\end{align}
The positive frequency mode represents a propagating wave with a
positive (negative) velocity in the direction of $y$ for $k_y>0$ ($k_y<0$).
The negative frequency mode is an unphysical mode and should be omitted.

Whereas, a TE mode satisfies
\begin{align}
 \frac{2\alpha}{i\omega \mu_0} - \frac{\sigma_0}{1-i\omega\tau}=0.
 \label{eq:te}
\end{align}
This equation has a unique solution exhibiting an overdamped oscillation,
\begin{align}
 \omega = - \frac{i}{\tau + \frac{\sigma_0 \mu_0}{2|k_y|}}.
 \label{eq:teomega}
\end{align}
The lifetime of the TE mode is enhanced in the long-wavelength limit $|k_y|\to 0$.
On the other hand, the equation does not admit an underdamped
oscillation with a non-zero real part of the frequency.

\begin{figure}[htbp]
 \begin{center}
  \includegraphics[scale=0.4]{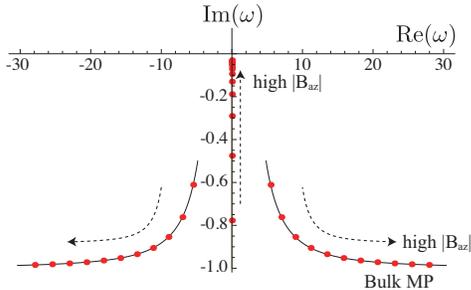}
 \end{center}
 \caption{(color online) The existence of three branches of
 magnetoplasmons is shown by flows with circles in the complex $\omega$-plane.
 The axes are expressed in units of ps$^{-1}$.
 The arrows denote the directions
 of the change in frequency as the magnetic field or $\omega_c$ increases
 from $10^{11}$ to $3\times 10^{13}$ s$^{-1}$. 
 We assumed that $\tau=1$ ps, $\alpha=1 \mu$m$^{-1}$, $\sigma_0=10^{-3}$
 $\Omega^{-1}$, and $\varepsilon=2.4$.
 The lifetime of the (bulk) magnetoplasmons in the high magnetic field
 limit corresponds to the value of $\tau$. 
 }
 \label{fig:empdisp}
\end{figure}

The system supports three eigenmodes in the presence of an external
magnetic field.
This is because we obtain the cubic equation with respect to
$\omega$, by making the determinant of the matrix of Eq.~(\ref{eq:mat1})
equal to zero, as
\begin{align}
 & \frac{4\epsilon}{\mu_0} \left\{ (1-i\omega \tau)^2 + (\omega_c \tau)^2
 \right\} \nn \\
 & -(1-i\omega \tau) \sigma_0 \left(\frac{2i\omega \epsilon}{\alpha} +
 \frac{2\alpha}{i\omega \mu_0} \right) + \sigma_0^2 = 0.
 \label{eq:deter}
\end{align}
The $\omega$ values of three solutions, which are calculated numerically, 
are shown in the complex $\omega$-plane in Fig.~\ref{fig:empdisp}
for the typical case of parameters.
We find that the system supports a purely relaxational state in the
presence of an external magnetic field.
The frequency is found analytically
when $\tau$ is sufficiently large ($|\omega_c| \tau \gg 1$)
as
\begin{align}
 \omega \simeq -\frac{i}{\tau + \frac{\omega_c^2}{\omega_p^2}\tau +
 \frac{\mu_0\sigma_0}{2\alpha}}.
 \label{eq:empseed}
\end{align}
The lifetime of this purely relaxational state is elongated by
increasing $|\omega_c|$ or in the long-wavelength limit.
The eigenvector is dominated by the TM component 
when $\sigma_0/\epsilon_0 c \ll (|\omega_c|\tau)^{3/2}$ is satisfied.
The dominance of the TM component is slightly peculiar because
Eq.~(\ref{eq:empseed}) reproduces the frequency of the TE mode
Eq.~(\ref{eq:teomega}) in the limit $\omega_c =0$.
The purely relaxational state is distinct from the bulk
magnetoplasmons with respect to the positions 
in the frequency domain and eigenvectors.
The dispersion relation of the bulk magnetoplasmons is obtained by
making the component that is proportional to $\epsilon\tau^2/\mu_0$
in Eq.~(\ref{eq:deter}) equal to zero as~\cite{Chiu1974}
\begin{align}
 \omega_{mp}(\omega_c) = \pm \sqrt{\omega_c^2 + \omega_p^2}-\frac{i}{\tau}.
 \label{eq:mp}
\end{align}
The negative frequency mode is an unphysical mode that should be omitted.
The eigenvectors of the modes are a hybrid of the TM and
TE modes. Generally, they are categorized as an elliptical polarization.
Practically, they are linear polarization because 
the TE component is dominant in a strong magnetic field ($\sigma_{xy}\gg \sigma_{xx}$).

It can be shown that the non-oscillating state found above starts
oscillating when the state is localized. 
We consider localized electric fields of the form,
\begin{align}
 \begin{split}
  & E_x = E_{x0} e^{i(k_y y -\omega t)} e^{-\beta x} e^{-\alpha|z|}, \\
  & E_y = E_{y0} e^{i(k_y y -\omega t)} e^{-\beta x} e^{-\alpha|z|}, \\
  & E_z = E_{z0} e^{i(k_y y -\omega t)} e^{-\beta x} e^{-\alpha|z|}.
 \end{split}
 \label{eq:Ebeta}
\end{align}
Here $\beta^{-1}$ is the lateral localization length from an edge and,
if $\beta=0$, 
the electric fields reproduce Eq.~(\ref{eq:Efield1}).
For the moment, we consider a positive $\beta$ by limiting our attention to
the bulk of the right half-plane of $x>0$.
Using the Maxwell equations
(see Appendix for detailed derivations),
we obtain a generalized version of Eq.~(\ref{eq:mat1}) with a modified
2$\times$2 matrix of the form: 
\begin{align}
 \begin{pmatrix}
  \frac{2\alpha}{i\omega \mu_0} \left( 1+ \frac{\beta^2}{\alpha^2} \right) - \sigma_{xx}(\omega)
  & -\frac{2k_y}{\omega \mu_0} \frac{\beta}{\alpha} - \sigma_{xy}(\omega) \cr
  -\frac{2k_y}{\omega \mu_0} \frac{\beta}{\alpha} + \sigma_{xy}(\omega)
  & \frac{2 i\omega \epsilon}{\alpha} - \sigma_{xx}(\omega) -
  \frac{2\beta}{i\omega \mu_0} \frac{\beta}{\alpha}
 \end{pmatrix}.
 \label{eq:mas_wa}
\end{align}
Each element of this matrix contains a term that is
proportional to $\omega^{-1}$, which 
is enhanced for the solution of Eq.~(\ref{eq:empseed}) 
because the $\omega$ is very close to the origin of the complex
$\omega$-plane.~\footnote{Although Eq.~(\ref{eq:empseed}) is a solution
obtained when $\beta=0$, we assume here that the corresponding state
exists when the localization is introduced. This is an assumption that
has been confirmed quickly by obtaining Eq.~(\ref{eq:dischiral}).}
This suggests that the eigenvector is modified accordingly so that 
the term is suppressed.
From Eq.~(\ref{eq:mas_wa}) the condition on the eigenvector is read off
as
\begin{align}
 \frac{\beta E_{y0} +ik_yE_{x0}}{i\omega}=0,
 \label{eq:constraint}
\end{align}
which is equivalent to an internal magnetic field $B_z$ being
suppressed in the bulk (see Eq.~(\ref{eq:Bz})).~\footnote{This is
consistent with the wide applicability of the theory of Volkov and
Mikhailov, in which the internal magnetic field is
neglected.~\cite{volkov88} 
The value of $\beta$ can be determined by using the frequency of
the magnetoplasmons (see Appendix for the details).}
By using Eq.~(\ref{eq:constraint}), 
we have the equations for the amplitudes as follows: 
\begin{widetext}
\begin{align}
 \begin{pmatrix}
  \frac{2\alpha}{i\omega \mu_0} \left( 1+ \frac{\beta^2}{\alpha^2} \right) - \sigma_{xx}(\omega)
  & -\frac{2k_y}{\omega \mu_0} \frac{\beta}{\alpha} - \sigma_{xy}(\omega) \cr
  \sigma_{xy}(\omega) & \frac{2 i\omega \epsilon}{\alpha} - \sigma_{xx}(\omega)
 \end{pmatrix}
 \begin{pmatrix}
  E_{x0} \cr E_{y0}
 \end{pmatrix}
 =0.
 \label{eq:mat4}
\end{align}
\end{widetext}
The vanishing determinant of the $2\times2$
matrix in Eq.~(\ref{eq:mat4}) leads to the following cubic equation with
respect to $\omega$:
\begin{align}
 & \omega_c^2 + \omega_p^2 -\omega^2 - \frac{i\omega \mu_0 \sigma_0}{2 \alpha
 (1 + \frac{\beta^2}{\alpha^2}) \tau^2} \nn \\
 & + \frac{\sigma_0 \mu_0}{2 \tau \alpha}
 \frac{\omega_p^2-\omega^2}{1+ \frac{\beta^2}{\alpha^2}} -
 \frac{\omega_p^2}{i\omega \tau} - \frac{\omega_c}{\omega}
 \frac{\omega_p^2 \beta k_y}{\alpha^2 + \beta^2} = 0.
 \label{eq:geq}
\end{align}
There is a solution that is approximated 
as 
\begin{align}
 \omega \simeq  \frac{\omega_p^2}{\omega_c}
 \frac{\beta}{\alpha^2 + \beta^2}k_y
 -\frac{i}{\tau + \frac{\omega_c^2}{\omega_p^2} \tau + \frac{\mu_0\sigma_0}{2\alpha(1+\frac{\beta^2}{\alpha^2})}} .
 \label{eq:dischiral}
\end{align}
The frequency has a non-zero real part, which is linear in $k_y$.
The fact that ${\rm Re}(\omega)/k_y \propto 1/\omega_c$ shows that the
mode is chiral (propagation direction is dependent on the sign of $B_{az}$).
Note also that the group velocity is suppressed when $|\omega_c|$ is
increased and the localization length $\beta^{-1}$ is fixed.
All these properties of this solution are consistent with those
of the edge magnetoplasmons 
and Eq.~(\ref{eq:dischiral}) can reproduce Eq.~(\ref{eq:empseed}) in
the $\beta=0$ limit.
Thus, we identify the mode with the edge magnetoplasmons.
It is worth noting that the intrinsic decay time of an edge magnetoplasmon
is derived from Eq.~(\ref{eq:dischiral}) as
\begin{align}
 \tau_{emp} \simeq \frac{\omega_c^2+\omega_p^2}{\omega_p^2}\tau,
 \label{eq:tauemp}
\end{align}
where $\mu_0\sigma_0/2\alpha(1+\frac{\beta^2}{\alpha^2})$ is omitted.
The ratio of $\tau_{emp}$ to $\tau$ increases as $|B_{az}|$ increases, 
and it is expressed in terms of the angular frequencies of the
magnetoplasmons as 
$\tau_{emp}/\tau=\omega^2_{mp}(\omega_c)/\omega^2_{mp}(0)$.
Note that $\tau$ can depend on $|B_{az}|$, and the dependence can be
determined from the lifetime of the magnetoplasmons (see Eq.~(\ref{eq:mp})).~\footnote{It is straightforward
to show that the other two solutions of Eq.~(\ref{eq:geq}) correspond to
the magnetoplasmons in the $\beta=0$ limit.}

\begin{figure}[htbp]
 \begin{center}
  \includegraphics[scale=0.45]{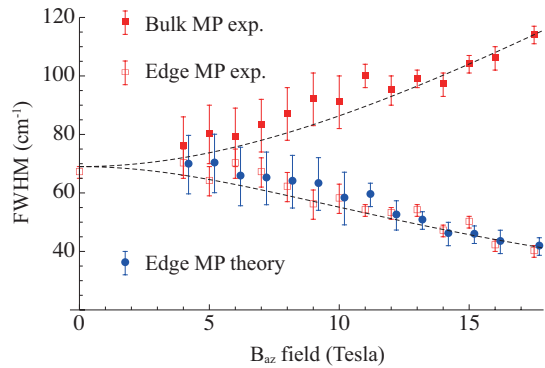}
 \end{center}
 \caption{(color online)  
 Equation~(\ref{eq:tauemp}) is applied to a recent experiment on the
 edge magnetoplasmons in graphene reported by Yan {\it et al.}~\cite{Yan2012}
 The close agreement between the circle and empty square plots
 supports the validity of Eq.~(\ref{eq:tauemp}). 
 The dashed curves are the plots of
 $69\sqrt{(\omega_c^2+\omega_p^2)/\omega_p^2}$ cm$^{-1}$
 and $69\sqrt{\omega_p^2/(\omega_c^2+\omega_p^2)}$ cm$^{-1}$.
 The errorbars with circles are shifted slightly horizontally from the
 proper values of the magnetic field in order to avoid overlap between
 the plots.
 }
 \label{fig:lifetime}
\end{figure}

Equation~(\ref{eq:tauemp}) can be used to explain a recent experiment on the
edge magnetoplasmons in graphene reported by Yan {\it et al.}~\cite{Yan2012}.
In Fig.~\ref{fig:lifetime}, magnetic field dependences of the full width
at half maximum (FWHM) of bulk and edge magnetoplasmons are indicated by
errorbars with filled and empty squares, respectively.
These are the experimental data taken from Fig.~2(D) in Ref.~\onlinecite{Yan2012}.
The plot of $2/c\tau_{emp}$ gives errorbars
with circles, where $\tau$ in Eq.~(\ref{eq:tauemp}) is taken from the experiment 
(the errorbars with filled squares).~\footnote{The $\omega_c$ value of
graphene is given by noting that the effective mass of the electrons
depends on the Fermi energy as $m=|E_F|/v^2$ ($v\simeq c/300$ is the
Fermi velocity of graphene). 
It is noted that by using the reported values (Fermi energy
$|E_F|=0.54$eV and average (relative) dielectric constant $\varepsilon=2.4$),
we obtain $\omega_p \simeq130$ cm$^{-1}$ by setting the wavevector $|{\bf k}|=0.4 \mu m^{-1}$.}
The close agreement between the positions of circles and empty squares
supports the validity of our result.
Moreover, $\tau_{emp}$ is elongated by decreasing $|k_y|$ through
$\omega_p$, which is consistent with a recent experiment on the
dissipation mechanism in graphene.~\cite{Kumada2014a}

Our derivation of the intrinsic lifetime of the edge magnetoplasmons
does not assume the details of the boundary of a 2DEG and therefore has
a wide application.
This is in contrast to the analyses of Volkov and Mikhailov, in which a
sharp electron density profile at the boundary (sharp edge) is assumed.
The edges of graphene and InAs meet this assumption, while GaAs might not
because of the existence of a depletion layer several micrometers thick.
Acoustic types of edge magnetoplasmons have been predicted for such
smooth edge.~\cite{Aleiner1994}
Note that since plasmon is the hybrid of electrons and electromagnetic
fields, it is difficult to identify the lifetime of edge magnetoplasmon
with the lifetime of electrons at only the edge channel.
Our formulation does not assume the specific properties of the electric
edge states, such as the absence of back scattering, but it yields
close agreement between Eq.~(\ref{eq:tauemp}) and the result in Ref.~\onlinecite{Yan2012}.
This fact suggests that because the experiment by Yan {\it et
al.}~\cite{Yan2012} is performed in the classical Hall effect region,
many electronic states including not only the edge
states but also (bulk) states near the edge (up to several micrometers
from the edge) are participating in the dynamics of the edge
magnetoplasmon.
The present model suggests an intriguing physical interpretation of the
longer lifetime based on the hybridization of the TE and TM modes.
A deviation from Eq.~(\ref{eq:tauemp}) may appear 
in the quantum Hall effect region and it can be attributed to the
contribution of the specific properties of the electric edge states.

\begin{figure}[htbp]
 \begin{center}
  \includegraphics[scale=0.6]{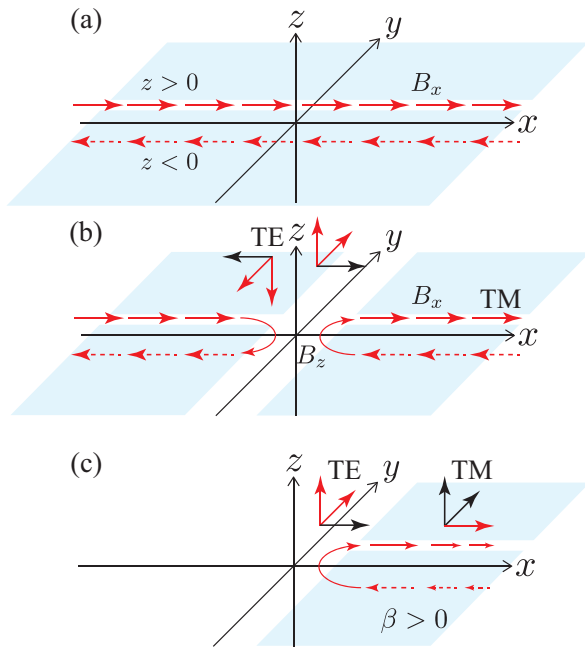}
 \end{center}
 \caption{(color online)
 (a) The magnetic field of a purely relaxational TM dominant state with
 $\beta=0$ is shown for $z>0$ and $z<0$.
 The sheet current $j_y \delta(z)$ that causes the discontinuity of
 $B_x$, is represented by a transparent layer.
 (b) When we introduce the edge along the $y$-axis, the magnetic fields at
 $z>0$ and $z<0$ must combine.
 The magnetic field at the edge has a non-zero $z$-component, which is a
 locally induced TE mode.
 (c) The edge-induced hybridization of the TE and TM modes results in
 an edge magnetoplasmon with a chiral propagation property.}
 \label{fig:hybrid}
\end{figure}

The finite real part of the frequency of an edge
magnetoplasmon originates from the mixing of $E_{x0}$ and $E_{y0}$. 
Namely, when we make a purely relaxational state localized by a
non-zero $\beta$, the eigenvector of the state is modified 
according to $\beta E_{y0} +ik_y E_{x0}=0$ for $x>0$.
An important feature of the modified eigenvector for $x<0$, for which
$\beta E_{y0} +ik_y E_{x0}=0$ does not need to be satisfied, can be
grasped pictorially without a mathematical calculation.
Suppose that as shown in Fig.~\ref{fig:hybrid}(a), 
a purely relaxational state, which is a TM dominant mode,
exists for a finite period of time in a periodic system without an edge.
Note that the magnetic field $B_x$ at $z=0+$ is pointing in the opposite
direction to that at $z=0-$. 
If we introduce the edge along the $y$-axis (by cutting the layer) in
Fig.~\ref{fig:hybrid}(b),
the magnetic fields at $z>0$ and $z<0$ must combine and form
a closed curve to satisfy one of the Maxwell equations: $\nabla \cdot {\bf B}=0$.
Thus, the magnetic field must have a non-zero $z$-component at the edge,
which is a locally induced TE mode.~\footnote{An extension of the work
by Volkov and Mikhailov~\cite{volkov88} gives a matrix Wiener-Hopf
equation, which is not solved in general. 
However, by using Eq.~(\ref{eq:constraint}) we can find that the
magnetic field $x<0$ is described by the modified Bessel function of the
second kind.}  
We may regard the edge magnetoplasmon as a composite of the TE mode
at the edge and the spatially decaying TM mode in the bulk, as shown in
Fig.~\ref{fig:hybrid}(c).

Since a TE mode is inevitably hybridized with the purely relaxational
state when the state changes into an edge magnetoplasmon, the fact that
a TE mode exhibits anomalous behavior in graphene is noteworthy. 
Mikhailov and Ziegler pointed out that the imaginary part of the dynamical
conductivity of graphene can be negative for a specific
frequency,~\cite{Mikhailov2007}
because an interband transition contributes to
the dynamical conductivity, while the Drude model only accounts for an
intraband transition.
As a result, they predict that graphene can support a TE mode for a
special frequency (even without an external magnetic field).
We can easily see from Eq.~(\ref{eq:te}) that 
an oscillating TE mode can appear when the imaginary part of the
dynamical conductivity is a negative number.
Bordag and Pirozhenko argued that the existence of an infinitesimal mass
gap in graphene leads to a special TE mode that propagates at the
speed of light.~\cite{Bordag2014}

In summary, the peculiarities of a purely relaxational state in the
bulk are partly eliminated by knowing their relationship to edge
magnetoplasmons, which have constituted the theme of various published
reports.~\cite{Allen1983,Glattli1985,Grodnensky1991,Ashoori1992,Tonouchi1994,Yan2012,Petkovic2013,Kumada2013}  
The strange behavior found for the state, 
such as the intrinsic long lifetime being proportional to the square of the
applied magnetic field, is our original conclusion that has not been
taken into account before.
Our result whereby the eigenvector is TM dominant with a very 
long-lived $E_z$ component has an advantage in that it detects the
signal via an electrode in a transient manner.~\footnote{The purely
relaxational state can exist in a gated sample.~\cite{Sasaki2014}
When a metal gate is placed on the dielectric media at a distance $d$
from the layer, the corresponding frequency can be calculated by
replacing $\epsilon$ and $\mu_0$ in Eq.~(\ref{eq:empseed}) with 
$\epsilon (1+\coth(\alpha d))/2$
and $\mu_0/(1+ \frac{\alpha_m + \alpha \tanh(\alpha d)}{\alpha +
\alpha_m \tanh(\alpha d)})$, respectively,
where $\alpha_m$ is the inverse of the localization length of a metal,
which may be taken to be $\infty$ at low frequencies.}
The local mixing with the TE mode may also imply that ferromagnetic
electrodes can excite the edge magnetoplasmons.

\section*{Acknowledgments}

K. S. is indebted to N. Kumada and H. Sumikura for discussions.

\appendix

\section{Derivation of Eq.~(\ref{eq:mat4})}

By applying Faraday's law to the electric fields Eq.~(\ref{eq:Ebeta}),
we obtain the magnetic fields for $z>0$ as
\begin{align}
 & i\omega B_x = ik_y E_z + \alpha E_y, \label{eq:Bx} \\
 & i\omega B_y = -\alpha E_x + \beta E_z, \label{eq:By} \\
 & i\omega B_z = -\beta E_y - ik_y E_x. \label{eq:Bz}
\end{align}
For $z<0$, the magnetic fields are given by replacing $\alpha$ with
$-\alpha$ and $E_z$ with $-E_z$ on the right-hand side.
As a result, $B_x$ and $B_y$ change their signs at $z=0$.
Note that the sign change of $E_z$ imposed at $z=0$ 
is still consistent with Gauss's law, which gives
\begin{align}
 -\beta E_x + ik_y E_y -\alpha E_z=0.
 \label{eq:Gauss}
\end{align}
It is useful to write Eqs.~(\ref{eq:Bx}), (\ref{eq:By}), and
(\ref{eq:Bz}) in the form of a $3\times 3$ matrix
as $i\omega {\bf B} = M {\bf E}$ where ${\bf E}={}^t(E_x,E_y,E_z)$,
${\bf B}={}^t(B_x,B_y,B_z)$, and 
\begin{align}
 M = 
 \begin{pmatrix}
  0 & \alpha & ik_y \cr
  -\alpha & 0 & \beta \cr
  -ik_y & -\beta & 0 
 \end{pmatrix}.
\end{align}
By applying Amp\'ere's circuital law for free space ($z \ne 0$)
where the charged current is absent (${\bf j}=0$),
$c^2 \nabla \times {\bf B} = \varepsilon \partial {\bf E}/\partial t$ , 
we obtain
\begin{align}
 \left( \frac{-i\omega \varepsilon}{c^2}\right) {\bf E} = M {\bf B}.
 \label{eq:MB}
\end{align}
By multiplying $i\omega$ with both sides and using 
$i\omega {\bf B} = M {\bf E}$, we have
$[\varepsilon \frac{\omega^2}{c^2}-M^2]{\bf E}=0$.
Thus, a non-vanishing electric field is possible when 
${\rm det}[\varepsilon \frac{\omega^2}{c^2}-M^2]=0$, namely, 
when 
\begin{align}
 \varepsilon \omega^2 \left( -k_y^2 + \alpha^2 + \beta^2 + \frac{\varepsilon \omega^2}{c^2}
 \right)^2 = 0
 \label{eq:det}
\end{align}
is satisfied.

The boundary conditions for the magnetic fields $B_x$ and $B_y$ at $z=0$
are expressed by 
\begin{align}
 & c^2 \left( B_{y}(z=0_-)-B_{y}(z=0_+) \right) = \frac{j_x(x,y)}{\epsilon_0}, \\
 & c^2 \left( B_{x}(z=0_+)-B_{x}(z=0_-) \right) =
 \frac{j_y(x,y)}{\epsilon_0}.
 \label{eq:con-2}
\end{align}
Putting Eqs.~(\ref{eq:Bx}) and (\ref{eq:By}) into these boundary
conditions and using Eqs.~(\ref{eq:Bz}), (\ref{eq:Gauss}), and
(\ref{eq:det}), we obtain Eq.~(\ref{eq:mas_wa}) or
\begin{widetext}
\begin{align}
 \begin{pmatrix}
  \frac{2\alpha}{i\omega \mu_0} \left( 1+ \frac{\beta^2}{\alpha^2} \right) - \sigma_{xx}(\omega)
  & -\frac{2k_y}{\omega \mu_0} \frac{\beta}{\alpha} - \sigma_{xy}(\omega) \cr
  -\frac{2k_y}{\omega \mu_0} \frac{\beta}{\alpha} + \sigma_{xy}(\omega)
  & \frac{2 i\omega \epsilon}{\alpha} - \sigma_{xx}(\omega) -
  \frac{2\beta}{i\omega \mu_0} \frac{\beta}{\alpha}
 \end{pmatrix}
 \begin{pmatrix}
  E_{x0} \cr E_{y0}
 \end{pmatrix}
 =0.
 \label{eq:mas}
\end{align}
\end{widetext}
This equation reproduces Eq.~(\ref{eq:mat4}) when 
\begin{align}
 \frac{\beta E_{y0} + ik_y E_{x0}}{i \omega}=0.
 \label{eq:bc}
\end{align}
It is noted that the determinant of the matrix of Eq.~(\ref{eq:mas})
is independent of the variable $\beta$.
Indeed, by setting $\beta=-ik_x$ in Eq.~(\ref{eq:Ebeta}), it is easily
understood that the inclusion of $\beta$ merely changes the propagation
direction. 
Thus, the solutions are given by Eqs.~(\ref{eq:empseed}) and
(\ref{eq:mp}), and therefore Eq.~(\ref{eq:mas}) fails to reproduces the edge
magnetoplasmons.
This clarifies the importance of the condition Eq.~(\ref{eq:bc}).

It is possible to estimate the value of $\beta$ using Eq.~(\ref{eq:bc}).
By putting the eigenstate of Eq.~(\ref{eq:mat4}) on Eq.~(\ref{eq:bc}),
we obtain 
\begin{align}
 \beta \simeq \frac{k_y}{\omega_c} \left( 1+
 \frac{\omega_c^2}{\omega_{p}^2}\right) \omega + i \frac{k_y}{\omega_c
 \tau},
 \label{eq:beta-app}
\end{align}
which is correct up to the first order of $\omega$.
By substituting Eq.~(\ref{eq:dischiral}) for $\omega$, we obtain
\begin{align}
 \alpha^2 + \beta^2 \simeq \frac{\omega_c^2 + \omega_p^2}{\omega_c^2} k_y^2.
\end{align} 
Thus, $\beta$ is determined as a function of $\alpha$.
We note that the $\alpha$ value is determined from $\omega_{p}$ or
$\omega_{mp}(\omega_c)$.
It is also worth noting that with Eq.~(\ref{eq:beta-app}), 
we can obtain $\omega$ as a function of $\alpha$
by eliminating $\beta$ from Eq.~(\ref{eq:det}).
The calculated $\omega$ reproduces 
Eq.~(\ref{eq:dischiral}).

\bibliographystyle{apsrev4-1}
%

%

\end{document}